\documentclass{article}
\usepackage{amssymb}

\input{tcilatex}

\begin{document}

\title{\textbf{INTERACTING VECTOR}$\ $\textbf{BOSON\ MODEL\ ANALYSIS\ OF}$\ $%
\textbf{THE }$^{160}$\textbf{Dy EXCITED\ STATES\ SPECTRUM}}
\author{{\large V.P. Garistov}$^{1,2}${\large , V. V. Burov}$^{2}$, \and 
{\large A.A. Solnyshkin}$^{2}$ \\
$^{1}${\small Institute for Nuclear Research and Nuclear Energy, BAS,
Bulgaria}\\
$^{2}${\small Joint Institute for Nuclear Research,Dubna, Russia} }
\maketitle

\begin{abstract}
Analysis of the recently obtained experimental data for collective states of 
$^{160}Dy$ is presented. The Interacting Vector Boson Model (IVBM) was
applied for the classification of low lying states with positive parity $%
0^{+},2^{+},4^{+},6^{+}$ and for description of rotational $ground$, $S$, $%
\gamma $ and two $octupole$ bands. The energies of the bands are reproduced
with high accuracy using only one set of the model parameters for all bands.
\end{abstract}

Independently on that the $^{160}Dy$ has a very \ complicated spectrum of
excited states, by now this nucleus is widely investigated experimentally.
During some last decades a great amount of experiments for the investigation
of the excited states spectrum of $^{160}Dy$ had been performed using
different types nuclear reactions [1-18], Coulomb excitation [19-25] and $%
\beta $ decay $^{160}Tb\rightarrow $ $^{160}Dy$ [26,27]$,$ $%
^{160m,g}Ho\rightarrow $ $^{160}Dy$ and $^{160}Er\rightarrow $ $%
^{160m,g}Ho\rightarrow $ $^{160}Dy$ [28-33]. The results of all the
investigations performed to 1996 year were analyzed in detail and presented
in Nuclear Data Sheets [34] and Table of Isotopes [35]. During some last
years using modern experimental techniques the comprehensive study of the $%
^{160}Er\rightarrow $ $^{160m,g}Ho\rightarrow $ $^{160}Dy$ $\beta $ decay
had been repeated with the measuring of $\gamma $ rays , conversional
electrons and $\gamma \gamma t$ coincidences spectra [36].These
investigations had supplemented the spectrum of the excited states of $%
^{160}Dy$ with more than 100 states and more than 500 $\gamma $ transitions
,and also to make agree with the data obtained from nuclear reactions and $%
\beta $ decay. At the same time a new study of excited states spectrum in $%
^{160}Dy$ using $^{7}Li$ ions beam with $^{158}Gd$ as a target had been
performed [37]. In these reactions were observed the ground band states with 
$K^{\pi }=0^{+}$ up to the excitation energy of $7231keV$ and $I^{\pi
}=28^{+}$, $\gamma $ band with $K^{\pi }=2^{+}$ to energy $6642keV$ and $%
I^{\pi }=25^{+}$, $S$ band to $4875keV$ with $I^{\pi }=20^{+}$ and octupole
bands $K^{\pi }=2^{-}$ to energy $6967keV$ with $I^{\pi }=26^{-}$\ and $%
K^{\pi }=1^{-}$ to energy $4937keV$ with $I^{\pi }=19^{-}$. 16 different
nature rotational bands in $^{160}Dy$ nucleus are identified. \ All this
made the $^{160}Dy$ nucleus a very good target for the theoretical nuclear
structure models. Recently these bands were analyzed [38] applying
Bohr-Mottelson model [39], $Q$ - phonon model [40], variable moment of
inertia model with dynamical asymmetry [41], Bohr-Mottelson model with \
Coriolis interaction [42]. In the same paper the positive parity states were
analyzed within the framework of $IBM-1$ [43].Using above approaches a
relatively good description for the states energies and transition
probabilities for low values of spins was obtained, but in the region of
higher values of spins the disagreement with experiment increased
noticeably. As a continuation of our theoretical analysis of very rich
experimental data for $^{160}Dy$ we apply the recently developed Interacting
Vector Boson Model ($IVBM$) [44]\ In this model within the framework of the
boson representation of the $sp(12,R)$ algebra all possible irreducible
representations of the group $SU(3)$ are determined uniquely through all
possible sets of the eigenvalues of the Hermitian operators $N$, $T^{2}$,
and $T_{0}$ or the equivalent $(\lambda ,\mu )$ labels in final reduction to
the $SO(3)$ representations, which define the angular momentum $L$ and its
projection $M.$

This model had illustrated a good description of the energies of different
rotational bands for the states with positive and negative parity for
instance in $^{226}Ra$ [44].

The detail description of this algebraic model one may find in [45]. Here we
present only necessary for our purposes expressions for energy spectrum and
the decomposition rules for the considering in the model chains (\ref{chains}%
).

\begin{equation}
\begin{tabular}{lllll}
$sp(12,R)$ & $\supset $ & $sp(4,R)$ & $\otimes $ & $so(3)$ \\ 
$\cup $ &  & $\cup $ &  & $\cap $ \\ 
$u(6)$ & $\supset $ & $u(2)$ & $\otimes $ & $su(3)$%
\end{tabular}
\label{chains}
\end{equation}

Written in terms of the $(\lambda ,\mu )$ labels facilitates together with
the decomposition rules \ the energy spectrum produced by the IVBM
Hamiltonian are as follow: 
\begin{eqnarray}
E((\lambda ,\mu );L;T_{0}) &=&\alpha N+\alpha _{1}N(N+5)+\beta _{3}L(L+1)+
\label{Spectr} \\
&&\alpha _{3}(\lambda ^{2}+\mu ^{2}+\lambda \mu +3\lambda +3\mu )+cT_{0}^{2}
\nonumber
\end{eqnarray}

\[
\ N-even\ \longrightarrow 0,2,4,6...... 
\]%
\ \ \ 
\begin{eqnarray*}
T &=&\frac{N}{2},\frac{N}{2}-1,\frac{N}{2}-2,...,0\ or\ 1 \\
\ T_{0} &=&-T,\ -T+1,...T
\end{eqnarray*}%
\[
\lambda =2T 
\]%
\begin{equation}
\mu =\frac{N}{2}-T  \label{NTcon1}
\end{equation}%
\[
K\ =\ min(\lambda ,\mu ),\ min(\lambda ,\mu )\ -2,.....\ ,0\ or\ 1 
\]%
\ \ 

$\qquad 
\begin{tabular}{llllll}
$K\ =\ 0\ \ \ \longrightarrow $ & $L=max(\lambda ,\mu ),$ & $L=max(\lambda
,\mu )\ -2$ & $...,$ & $0,$ & $1$ \\ 
$K\neq \ 0\ \ \ \longrightarrow $ & $L=max(\lambda ,\mu ),$ & $L=max(\lambda
,\mu )-1,$ & $...,$ & $0,$ & $1$%
\end{tabular}%
$

The parity of the states is defined as $\pi =(-1)^{T}$. The index $K$
appearing in this reduction is related to the projection of $L$ in the body
fixed frame and is used with the parity to label the different bands in the
energy spectra of the nuclei. This allows us to describe both positive and
negative parity bands. Further we use the connection:

\begin{equation}
N=4L
\end{equation}

Thus, taking into account the reducing rules (\ref{NTcon1}) the energy can
be rewritten only in terms of the pseudospin $T$ and angular momentum $L$ :

\begin{equation}
\begin{tabular}{l}
$E(L,T)=4\,\alpha \,L+4\alpha _{1}\,L\,\left( 5+4\,L\right) \,+\beta
_{3}L(L+1)+\,$ \\ 
\\ 
$\ \ \ \ \ \ \ \ \ \ \ \ \ \ \ \ \alpha _{3}\left[ 6\,L+4\,L^{2}\,+3\,T^{2}%
\,+3T\right] +c\,T_{0}^{2}$ .%
\end{tabular}%
\end{equation}

In our investigations of the experimental excited states in $^{160}Dy$ we
start with the study of the low lying $0^{+}$ states within the framework of
simplified pairing vibrational model.\ With other words we make the
classification of $0^{+}$ states basing on phenomenological monopole part of
\ collective Hamiltonian for single level approach written in terms of boson
creation and annihilation operators\ \ $R_{+}$ , $R_{-}$ and $R_{0}$ 
\begin{equation}
H=\alpha R_{+}R_{-}+\beta R_{0}R_{0}+\frac{\beta \Omega }{2}R_{0}{\large ,\ }
\label{H1}
\end{equation}%
constructed with the pairs of fermion operators $a^{\dagger }$ and $a$ of
the fermions placed at subshell $j$ and\ model\ parameters$\ \alpha $,$\beta 
$\textbf{\ }\ and$\ \Omega =\frac{2j+1}{2}$: 
\begin{equation}
\begin{tabular}{l}
$R_{+}={\frac{1}{2}}\sum\limits_{m}(-1)^{j-m}\alpha _{jm}^{\dagger }\alpha
_{j-m}^{\dagger }\;$, \\ 
$R_{-}={\frac{1}{2}}\sum\limits_{m}(-1)^{j-m}\alpha _{j-m}\alpha _{jm}\;,$
\\ 
$R_{0}={\frac{1}{4}}\sum\limits_{m}(\alpha _{jm}^{\dagger }\alpha
_{jm}-\alpha _{j-m}\alpha _{j-m}^{\dagger })\;.$ \\ 
$\left[ R_{0},R_{\pm }\right] =\pm R_{\pm },\ \ \ \ \ \ \ \left[ R_{+},R_{-}%
\right] =2R_{0}$%
\end{tabular}
\label{ROPER}
\end{equation}

\ Applying the Holstein-Primakoff [47] transformation to the operators $%
R_{+} $ , $R_{-}$ and $R_{0}$

\begin{equation}
\begin{array}{ccc}
R_{-}=\sqrt{2\Omega -b^{\dagger }b}\;b\text{; \ \ \ } & R_{+}=b^{\dagger }%
\sqrt{2\Omega -b^{\dagger }b}\text{; \ \ \ } & R_{0}=b^{+}b-\Omega .%
\end{array}%
\end{equation}

where $b^{\dagger }$, $b$ are new ( ideal ) boson creation and annihilation
operators with commutation rules:

\begin{equation}
\left[ b,b^{\dagger }\right] =1\,,\;\left[ b,b\right] =\left[ b^{\dagger
},b^{\dagger }\right] =0.\ \ \ \   \label{boson}
\end{equation}

Now, the initial Hamiltonian (\ref{H1}) \ written in terms of ideal bosons
has the form:

\begin{equation}
H=Ab^{\dagger }b-Bb^{\dagger }bb^{\dagger }b.  \label{Hampure}
\end{equation}

\[
A=\alpha (2\Omega +1)-\beta \Omega ,\,\,B=\ \alpha -\beta . 
\]%
$\ \ \left\vert n\right\rangle $ - boson state is determined as: 
\begin{equation}
\ \ \left\vert n\right\rangle =\frac{1}{\sqrt{n!}}(b^{+})^{n}\left\vert
0\right\rangle ,\text{where}\ \ b\left\vert 0\right\rangle =0\;\ 
\end{equation}

Thus the energy spectrum produced by Hamiltonian (\ref{Hampure}) \ is the
parabolic function of the number of ideal monopole \ bosons $n$: $\ \ \ \ \
\ \ \ \ \ \ \ \ \ \ \ \ \ \ \ \ \ $%
\begin{equation}
E_{n}=An-Bn^{2}  \label{E(n)}
\end{equation}%
In Figure 1 we show the new representation of available experimental data
for $0^{+}$ states in $^{160}Dy$ as distributed by number of ideal monopole
bosons (\ref{boson}) The average energy deviations $<|E_{expt}-E_{calc}|\ >$
are $10.7$ $KeV$. The parameters $\ A$ and $B$ of (\ref{E(n)}) are evaluated
by fitting the experimental energies of the different $0^{+}$ states of a
given nucleus to the theoretical ones applying all possible permutations of
the classification numbers $n$ and extracting the distribution corresponding
to the minimal value of $\chi $ - square. In Figure 2 as an additional
example we present our results for description of $0^{+}$ excited states in
new experimental data for $^{158}Gd$ nucleus. Even in this case we obtain a
very good agreement between our distribution and experiment - the average
deviation less than $13\ KeV$ per point. With nice accuracy the experimental
energies for low lying collective states follow the parabolic distribution
function of number of collective excitations. Now we can label every $K^{\pi
}=0^{+}$\ state by an additional characteristic $n$ number of monopole
bosons determining it's collective structure. It is interesting to point out
that the ordering of the states in respect to their number of phonons does
not necessarily correspond to increase of excitation energy. For some nuclei
the lowest excited $K^{\pi }=0^{+}$ states have more collective structure (
lager $n$ ) than the states with higher excitation energies.

\ Of course it is straightforward now to see whether the low lying excited
states having different from zero spin can be also represented in the same
form of the energies distributed by parabolic type function and can we
consider the new classification parameter as a measure of collectivity
determining each low lying state.

Using the $(\lambda ,\mu )$ labels facilitates and choosing for instance $%
(\lambda ,0)$ multiplet \ together with the reducing rules (\ref{NTcon1})
after simple regrouping of the terms in (\ref{Spectr}) for any fixed value
of angular momentum $L$ the energy spectrum corresponding to this $(\lambda
,0)$ multiplet is:

\begin{equation}
E(\lambda )=A\lambda -B\lambda ^{2}+C  \label{Elam}
\end{equation}%
here $A$, $B$ and $C$ are the combinations of free model parameters of (\ref%
{Spectr}) $\alpha $, $\alpha _{3}$, $\beta _{3}$ and $\alpha _{1}$. Hence
choosing any permitted by (\ref{NTcon1}) $(\lambda ,\mu )$ multiplet\ we
again may classify the low lying excited states energies in even even nuclei
applying the parabolic type distribution function and considering label $C$
as a measure of collectivity of the corresponding excited states. In Figure
3 we show the distributions of the $2^{+}$, $4^{+}$, $6^{+}$ excited states
in $^{160}Dy$. In these figures we use the notion $n=\frac{\lambda }{4}$.
And in this case we have a good description with the average energy
deviations $<|E_{expt}-E_{calc}|\ >$ $39.4$, $21.9$, and $50.4\ keV$ for the
levels with $I^{\pi }=$ $2^{+}$, $4^{+}$, and $6^{+}$ respectively.

From the energy spectrum written in terms of $\lambda $ and $\mu $ labels (%
\ref{Spectr}) we can determine\ the expressions for the rotational bands
energies as follow:

$K^{\pi }=0^{+}$ ground-state band $\{\lambda =0,\mu =2L\}$ and $T=0$, $%
T_{0}=0$

\begin{equation}
E_{gr}=4\alpha L+\beta _{3}L(L+1)+4\alpha _{1}L(5+4L)+2\alpha _{3}(6L+4L^{2})
\label{grband}
\end{equation}

$K^{\pi }=0^{+}$ $S$ band $\{\lambda =4,\mu =2L-2\}$ and $T=2$, $T_{0}=1$

\begin{equation}
E_{s}=c_{1}+4\alpha L+\beta _{3}L(L+1)+4\alpha _{1}L(5+4L)+2\alpha
_{3}(6L+4L^{2})  \label{Sband}
\end{equation}

$K^{\pi }=1^{-}$ octupole vibrational band $\{\lambda =2,\mu =2L-1\}$ and $%
T=1$, $T_{0}=1,L\geq 1$

\begin{equation}
E_{oct~1^{-}}=c_{1^{-}}+4\alpha L+\beta _{3}L(L+1)+4\alpha
_{1}L(5+4L)+\alpha _{3}\{10+5(2L-1)+(2L-1)^{2}\}  \label{foctband}
\end{equation}

$K^{\pi }=2^{+}$ $\gamma $ vibrational band $\{\lambda =4,\mu =2L-2\}$ and $%
T=2$, $T_{0}=1,L\geq 2$

\begin{equation}
E_{\gamma }=c_{2^{+}}+4\alpha L+\beta _{3}L(L+1)+4\alpha _{1}L(5+4L)+\alpha
_{3}\{28+7(2L-2)+(2L-2)^{2}\}  \label{gammaband}
\end{equation}

$K^{\pi }=2^{-}$ octupole vibrational band $\{\lambda =2,\mu =2L-1\}$ and $%
T=1$, $T_{0}=1,L\geq 2$

\begin{equation}
E_{oct\ 2^{-}}=c_{2^{-}}+4\alpha L+\beta _{3}L(L+1)+4\alpha
_{1}L(5+4L)+\alpha _{3}\{10+5(2L-1)+(2L-1)^{2}\}  \label{secoctband}
\end{equation}

Our previous calculations of rotational bands energies with different forms
of nuclear density shapes [46] had shown that the moment of inertia depends
on number of monopole bosons $n$ approximately as:

\begin{equation}
I(n)\thickapprox I(0)(1+xn)
\end{equation}%
where $x$ is connected with the diffuseness parameter $s,$ compressibility
coefficient ${C_{0}}$, one-phonon energy $E_{0}\,$ {and nuclear half-radius }%
$R${\ as:}

\begin{equation}
x=\frac{E{_{0}}\,R^{2}\,\left( \left( -3+20\,\pi \right) \,R^{4}+30\,\left(
-1+4\,\pi \right) \,R^{2}\,s^{2}+45\,\left( -1+4\,\pi \right) \,s^{4}\right) 
}{8\,C{_{0}}\,{\pi }^{2}\,\left(
R^{6}+13\,R^{4}\,s^{2}+45\,R^{2}\,s^{4}+45\,s^{6}\right) }
\end{equation}

\bigskip That is why in our calculations we choose parameter $\beta _{3}$ to
be 
\begin{equation}
\beta _{3}=\frac{1}{2I(n)}=\frac{\beta _{0}}{1+nx}  \label{beta}
\end{equation}

We apply this approximation in our calculations of the energies of
rotational bands determined by (\ref{grband}-\ref{secoctband}). In Figures 4
- 5. \ is shown the comparison of our calculations with experiment. It is
important to point out that all these bands are calculated with the same set
of parameters : 
\begin{equation}
\begin{tabular}{|c|c|c|c|c|}
\hline
$\alpha $ & $\alpha _{1}$ & $\alpha _{3}$ & $x$ & $\beta _{0}$ \\ \hline
$\ 0.00568396$ & $-0.0274276$ & $0.05485516$ & $\ \ 0.0020997$ & $0.23$ \\ 
\hline
\end{tabular}
\label{parameters}
\end{equation}

In these figures are also presented the corresponding values of number of
monopole bosons $n$ building the collective \ excited $0^{+}$ state
(entering ( \ref{beta})) which mainly determines the moment of inertia of \
each band. The agreement between calculated and experimental energies is
very good and average energy deviation $<|E_{expt}-E_{calc}|\ >$ for all
bands under consideration is less than $9$ $KeV$ per point.

For vindication of the right positions of the bands with different parities
and check the parity splitting we had calculated the staggering functions of
the fifth order for experimental points and calculated data:\medskip

\begin{equation}
\begin{tabular}{c}
$\Delta ^{5}E(L)=6(E(L)-E(L-1))-4(E(L-1)-E(L-2))-$ \\ 
\\ 
$4(E(L+1)-E(L))+E(L+2)-E(L+1)+E(L-2)-E(L-3)$%
\end{tabular}
\label{delta}
\end{equation}

According to our calculations for the energies of the bands (Figure 4
-Figure 5) we had calculated the staggering functions with replaced states $%
I^{\pi }=18^{+},4.181$, $MeV$, $I^{\pi }=20^{+},4.875\ MeV$\ from $S$ band
to the ground band while the states \ $I^{\pi }=18^{+}$, $3.67$ $MeV$, $%
I^{\pi }=20^{+},4.279$, $I^{\pi }=22^{+},4.936$ $MeV$, \ $I^{\pi
}=24^{+},5.648$ $MeV$, $I^{\pi }=26^{+},6.413$ $MeV$, $I^{\pi }=28^{+},7.231$
$MeV$ \ from the ground to $S$ band (which really produces much better
agreement with experiment than the calculations with previous straightening
[37] . Hence we have proposed that the sequence of states $\ $ $I^{\pi
}=18^{+},4.181$, $MeV$, $I^{\pi }=20^{+},4.875$ $MeV$ belongs to the ground
band while the states $I^{\pi }=18^{+}$, $3.67$ $MeV$, $I^{\pi
}=20^{+},4.279\ MeV$, $I^{\pi }=22^{+},4.936$ $MeV$, \ $I^{\pi
}=24^{+},5.648 $ $MeV$, $I^{\pi }=26^{+},6.413$ $MeV$, $I^{\pi
}=28^{+},7.231 $ $MeV$ must be related to the $S$ band, moreover that for
simultaneous description of the bands with previous straightening [37] the
additional parameter is required. In Figure 7 are presented odd-even
staggering functions ( \ref{delta}) for $S$ ($K^{\pi }=0^{+}$) and octupole (%
$K^{\pi }=2^{-}$) bands, $S $ ($K^{\pi }=0^{+}$) and $\gamma $ ($K^{\pi
}=2^{+}$) bands, for $S$ ($K^{\pi }=0^{+}$)and octupole $S$ ($K^{\pi }=1^{-}$%
) bands, ground ($K^{\pi }=0^{+}$) and $\gamma $ ($K^{\pi }=2^{+}$) bands
calculated for experimental and theoretical data. The agreement between
theory and experiment is good. In Figure 8 we show the comparison of the
staggering functions for octupole \ band ($K^{\pi }=1^{-}$) and the band \ ($%
K^{\pi }=0^{+}$), so far determined as a ground band in [37]. In the case of
straightening [37] the agreement of the calculated and experimental data is
sensitively worse. The same situation one can find in Figure 9, where are
compared staggering functions for octupole ($K^{\pi }=2^{-}$) \ and ground
bands. To prove that this rearrangement of some of states between $S$ and
ground bands is not a sort of mere assertion we must analyze the behavior of
the $B(E2)$ transitions in the region of crossing ground and $S$ bands
states. Indeed, the transition probability even in simple rigit rotor model
depends on intrinsic quadrupole momentum\ $Q_{0}$ that in our consideration
is a function of number of monopole bosons and increases with \ increase of
number of monopole bosons $n$ [44]. Thus it should be good to make detail
theoretical analysis of $B(E2)$ transitions within the framework of our
model. This work is in progress.

The investigation was supported in part by the RFBR and \ by Bulgarian
Science Foundation under contract $\Phi $ 905.

\end{document}